\documentclass[11pt,letterpaper]{article}
\usepackage{fullpage,amsmath,amsthm,amsfonts} 
\usepackage{graphicx}

\newcommand{\ket}[1]{\lvert #1 \rangle}
\newcommand{\bra}[1]{\langle #1 \rvert}
\newcommand{\braket}[2]{\langle #1 | #2 \rangle}

\newcommand{\size}[1]{\lvert #1 \rvert}
\newcommand{\abs}[1]{\lvert #1 \rvert}
\newcommand{\length}[1]{|{#1}|}
\newcommand{\given}{|}

\newcommand{\eps}{\varepsilon}
\renewcommand{\phi}{\varphi}

\newcommand{\norm}[1]{\parallel #1 \parallel}

\newcommand{\leftload}{\mathbf{LL}}
\newcommand{\rightload}{\mathbf{RL}}

\newcommand{\qqc}{\mathsf{QQC}}
\newcommand{\rqc}{\mathsf{RQC}}
\newcommand{\K}{\mathsf{K}}
\newcommand{\C}{\mathsf{C}}

\newtheorem{theorem}{Theorem}
\newtheorem{corollary}{Corollary}

\newtheorem{definition}{Definition}
\newtheorem{proposition}{Proposition}
\newtheorem{lemma}{Lemma}
\newtheorem{claim}{Claim}

\theoremstyle{definition}
\newtheorem{example}{Example}

\begin{document}

\author{Sophie Laplante\thanks{
LRI, UMR 8623 CNRS and Universit\'e Paris--Sud,
91405 Orsay, France,
emails:  \{{\tt laplante}, {\tt magniez}\}{\tt @lri.fr};
partially supported by
the EU 5th framework programs
RESQ IST-2001-37559 and RAND-APX IST-1999-14036,
and by ACI Cryptologie CR/02 02 0040
and ACI S\'ecurit\'e Informatique 03 511
grants of the French Research Ministry.}
\and
Fr\'ed\'eric Magniez${}^{*}$}

\title{Lower bounds for randomized and quantum query complexity\\
using Kolmogorov arguments}

\date{}

\maketitle

\begin{abstract}
We prove a very general lower bound technique 
for quantum and randomized query complexity, that is
easy to prove as well as to apply.  
To achieve this, we introduce the use of Kolmogorov 
complexity to query complexity.
Our technique generalizes the weighted, unweighted methods of Ambainis,
and the spectral method of Barnum, Saks and Szegedy.
As an immediate consequence of our main theorem, 
adversary methods can only prove lower bounds 
for  boolean functions $f$ in $O(\min(\sqrt{n C_0(f)},\sqrt{n C_1(f)}))$,
where $C_0, C_1$ is the certificate complexity,
and $n$ is the size of the input.
We also derive a general form of the {\it ad hoc} weighted method
used by H{\o}yer, Neerbek and Shi to give a quantum lower bound 
on ordered search and sorting.
\end{abstract}

\section{Introduction}
\subsection{Overview}

In this paper, we study lower bounds for randomized and quantum 
query complexity.
In the query model, the input is accessed using oracle queries,
and the query complexity of an algorithm is the number of calls to
the oracle.
Since it is difficult to obtain lower bounds on time directly,
the query model is often used to prove concrete lower bounds,
in classical as well as quantum computation.

The two main tools for proving lower bounds of
randomized query complexity,
the polynomial method~\cite{bbcmw01}
and the adversary method~\cite{amb02},
were successfully extended to quantum computation.
In the randomized setting, the adversary method is
most often applied using Yao's minimax principle~\cite{yao77}.
Using a different approach, which introduces the notion of
quantum adversaries,
Ambainis developed a general scheme in which it suffices to analyze
the combinatorial properties of the function in order to
obtain a quantum lower bound.
Recently, Aaronson~\cite{aar03} brought these combinatorial
properties back to randomized computation, using Yao's minimax principle.

The most general method for proving lower bounds in quantum query complexity
is the semidefinite programming method of Barnum, Saks and Szegedy~\cite{bss03}. 
This method is in fact an exact characterization of the query complexity.
However, the method is so general as to be very difficult to apply to obtain
concrete lower bounds.
Barnum, Saks and Szegedy gave a weaker method
derived from the semidefinite programming approach, 
using weight matrices and their largest eigenvalue.
This spectral method can be thought of as a generalization of Ambainis'
unweighted method.  
Other generalizations of Ambainis' unweighted method have been previously
introduced~\cite{bs02,amb03:vs}. 
All of them use a weight function on the instances.
The difficulty in applying these methods is finding a 
good weight function on the instances.
H{\o}yer, Neerbek and Shi~\cite{hns02} were the first to use such
weight assignments to prove lower bounds for 
searching in ordered lists and sorting. Their {\it ad hoc} method,
though similar in many respects, does not
fall into setting of the weighted method of Ambainis~\cite{amb03:vs}.

This paper presents a new, very general adversary technique
(\textbf{Theorem~\ref{thm:general}}) to prove lower bounds in
quantum and randomized query complexity. 
We believed that this technique is simpler to prove and to apply.
It is based on the framework of Kolmogorov complexity.
This framework has proven to be very useful for proving negative
results in other models of computation, for example for number of 
rounds and length of advice in random-self-reductions
in~\cite{ffln98,bl99}.   The techniques we use here are 
an adaptation of those techniques to the framework of 
query complexity.
We expect that this framework will not only
prove to be useful for negative results in other quantum models
of computation, for instance, communication complexity, but
also for finer analysis of query complexity in terms of the number 
of rounds of queries.

The proof of Theorem~\ref{thm:general} is in two parts.
The first part (\textbf{Divergence Lemma}) shows how fast the computations
can diverge when they start on different inputs.  This part
depends on the model of computation (randomized or quantum).
The quantum case of this lemma was first proven by Ambainis~\cite{amb02}.
The second part (\textbf{Query Information Lemma}) 
does not depend on the model of computation.
It establishes the relationship between 
the Kolmogorov complexity of individual positions of the input, 
and the probability that a given algorithm makes a query to this position.
Whereas Aaronson~\cite{aar03} used a different approach to prove a version
of Ambainis'  method for randomized algorithms, here we use the
same framework to establish lower bounds 
for both quantum and randomized query complexities ($\qqc$ and $\rqc$).

We show that our method encompasses all previous adversary methods,
including the quantum and randomized weighted methods~\cite{amb03:vs,aar03}
(\textbf{Theorem~\ref{thm:ambainis+weights}})
 and the spectral method~\cite{bss03} (\textbf{Theorem~\ref{thm:spectra}}).  
As an immediate consequence of our main theorem (observed by Troy Lee), 
our method can only prove lower bounds 
for  boolean functions in $O(\min(\sqrt{n C_0(f)},\sqrt{n C_1(f)}))$,
where $C_0$ and $C_1$ is the certificate complexity
of negative and positive instances, respectively,  of $f$,
and $n$ is the size of the input (\textbf{Theorem~\ref{thm:troy}}).
Prior to our work, it was known~\cite{amb03:vs} that 
the unweighted Ambainis method~\cite[Theorem~5.1]{amb02}
could not prove bounds better than $\Omega(\sqrt{C_0(f) C_1(f)})$;
Szegedy~\cite{sze03} also proved independently that the semidefinite
programming method could not prove lower bounds
better than $O(\min(\sqrt{n C_0(f)},\sqrt{n C_1(f)}))$,
and Zhang~\cite{zha03} proved the same thing for Ambainis' weighted method.

We also give a generalization 
(\textbf{Theorem~\ref{thm:hoyer}})
of the {\it ad hoc} proofs of
H{\o}yer, Neerbek and Shi~\cite{hns02} as a corollary of our method.
For this we introduce a new distance scheme.
This new scheme separates the quantum part from the combinatorial part
of these {\it ad hoc} proofs. Using it, we prove 
the lower bound of~\cite{hns02} using only combinatorial arguments.  
We end the paper by giving some applications of our method
to prove lower bounds for some graph properties: 
bipartiteness (\textbf{Theorem~\ref{thm:bipart}}) 
and
connectivity (\textbf{Theorem~\ref{thm:connect}}).
This lower bounds were proven in~\cite{dhhml03}.
We reprove them here to illustrate the simplicity of our method.

\newpage

\subsection{Main result} 
Our main result is stated below.  
\begin{theorem}
\label{thm:general}
There exists a constant $C>0$ such that the following holds.
Let $\Sigma$ be a finite set, let $n\geq 1$ be an integer,
and let $S\subseteq\Sigma^n$ and $S'$ be sets.
Let $f:S\rightarrow S'$.
Let $A$ be an algorithm that for all $x\in S$ computes $f$,
with bounded error $\eps$ and at most $T$ queries to the input.
Then for every $x,y\in S$ with $f(x)\neq f(y)$:
\begin{enumerate}
\item If $A$ is a quantum algorithm then
$$T \geq C\times
      \frac{1-2\sqrt{\eps(1-\eps)}}{\sum_{i:x_i \neq y_i}
                            \sqrt{2^{-\K(i \given x,A)-\K(i \given y,A)}}};$$
\item If $A$ is a randomized algorithm then
$$T \geq C\times
\frac{1-2\eps} 
{\sum_{i:x_i \neq y_i} 
\min\left(2^{-\K(i \given x,A)},2^{-\K(i \given y,A)}
\right)}.$$
\end{enumerate}
\end{theorem}

We briefly describe the intuition behind the proof of Theorem~\ref{thm:general}. 
Consider an algorithm that purports to compute  $f$,
presented with two inputs $x,y$ that lead to  different
outputs.  The algorithm must query those positions
where $x$ and $y$ differ with average probability of the order of $\tfrac{1}{T}$, 
or it will not successfully compute the function.  
On the other hand, the queries that are 
made with high average probability 
can be described succinctly given the input and the algorithm, 
using the Shannon-Fano code. If we exhibit
a pair of strings $x,y$ for which there is no succinct description
of any of the positions where $x$ and $y$ differ,
then the number of queries must be large.

The same reasoning can be applied
to classical and to quantum computing; the only difference is 
how fast two different input states cause the outputs to diverge
to different outcomes.

To conclude the introduction we give a very simple application,
for Grover search.
\begin{example}
Fix $n$ and a quantum algorithm $A$ for Grover search for instances of length $n$.
Let $z$ be a binary string of
length $\log n$, with $\K(z \given A) \geq \log n$.
Let $j$ be the integer between $0$ and $n-1$ whose binary expansion is  $z$.
Consider $x$, the all $0$'s string, and let $y$ be everywhere 
$0$ except at position $i=j+1$, where it is $1$.
Then $\K(i \given x,A) \geq \log n -O(1)$ and $\K(i \given y,A) = O(1)$,
therefore,
$\qqc(\textsc{Search})=\Omega(\sqrt{n})$.
\end{example}

\section{Preliminaries}
\subsection{Kolmogorov complexity}
We use a few standard results in Kolmogorov complexity
and information theory in this paper.  We briefly review these
here.  The reader is invited to consult standard textbooks such
as \cite{lv97} for more background
on Kolmogorov complexity, and \cite{ct91} for more on information theory.
We denote the length of a finite string $x$ by $\length{x}$.
We assume that the Turing machine's alphabet is the same finite alphabet
as the alphabet used to encode instances of the function under consideration.
Letters $x,y$ typically represent instances; $i$ is an index into the
representation of the instance;  and~$p,q$ are
probability distributions.  Programs are denoted $P$, and the output
of a Turing machine $M$ on input $x$ is written $M(x)$.  When there
are multiple inputs, we assume that a standard encoding of tuples is used.
\begin{definition}
Let $M$ be a Turing machine.
Let $x$ and $y$ be finite strings.
\begin{enumerate}
\item
The {\em Kolmogorov complexity of $x$ given $y$ with respect to $M$} is
denoted $\C_{M}(x \given y)$,  and defined as follows:
$$\C_{M}(x \given y)=
    \min (\mbox{$\length{P}$ such that $M(P,y)= x$}).$$
\item A set of strings is {\em prefix-free} if no string is a prefix of another 
in the set.
\item The {\em prefix-free Kolmogorov complexity of $x$ given $y$ 
with respect to $M$} is
denoted $\K_{M}(x \given y)$, and defined as follows:
$$\K_{M}(x \given y)=\min(\mbox{$\length{P}$ such that $M(P,y)= x$}),$$
where $P$ is taken in some fixed prefix-free set.
\end{enumerate}
\end{definition}
In the rest of the paper $M$ is some fixed universal Turing machine, and
we will write $\C$ and $\K$ instead of $\C_M$ and $\K_M$.
When $y$ is the empty string, we write $\K(x)$ instead of $\K(x\given y)$.

\begin{proposition}
\label{prop:knowing}
There exists a constant $c\geq 0$ such that
for every finite string $\sigma$,
$$\K(x \given \sigma)\leq \K(x)+c, \text{ and }$$
$$ \K(x) \leq \K(\sigma) + K(x \given \sigma) + c.$$
\end{proposition}

\begin{proposition}[Kraft's inequality]\label{prop:kraft}
Let $S$ be any prefix-free set of finite strings.
Then $\sum_{x\in S}2^{-\abs{x}} \leq 1.$
\end{proposition}


\begin{proposition}[Shannon's coding theorem]\label{prop:shannon}
Consider a source~$\mathcal{S}$ of finite strings
where $x$ occurs with probability $p(x)$. Then
for any code for $\mathcal{S}$, the average code length is bounded
below by the entropy of the source, that is,
if $x$ is encoded by the code word $c(x)$
of length $\size{c(x)}$,
$H(\mathcal{S}) = \sum_{x:p(x)\neq 0}p(x)\log(\frac{1}{p(x)}) 
\leq \sum_{x:p(x)\neq 0}p(x)\abs{c(x)}$.
\end{proposition}

\begin{lemma}
\label{lemma:shannon-incompressibility}
Let $\mathcal{S}$ be a source as above.  Then for any fixed finite string $\sigma$,
there exists a string $x$ such that $p(x)\neq 0$
and $\K(x \given \sigma) \geq \log(\frac{1}{p(x)})$.
\end{lemma}
\begin{proof}
By Shannon's coding theorem, 
$$H(\mathcal{S}) = \sum_{x:p(x)\neq 0}p(x) \log(\tfrac{1}{ p(x)}) 
                \leq \sum_{x:p(x)\neq 0}p(x) \K(x \given \sigma),$$
because $\K(x \given \sigma)$ is the length of an encoding of
$x$.  
Therefore there exists  $x$ such that $p(x)\neq 0$ and
$\K(x) \geq \log(\frac{1}{p(x)}) .$
\end{proof}

The Shannon-Fano code is a prefix-free code that
encodes each word $x$ with $p(x)\neq 0$, using
$\lceil\log(\frac{1}{p(x)})\rceil$ bits.  
We will write $\log(\frac{1}{p(x)})$ to simplify notation.
The code can easily be
computed given a description of the probability distribution.
This allows us to write the following proposition, where
$\K(x \given \mathcal{S})$ means the prefix-free Kolmogorov complexity of $x$
given a finite description of $\mathcal{S}$.
\begin{proposition}[Shannon-Fano code]
There exists a constant $c\geq 0$, such that for
every source~$\mathcal{S}$ as above,
for all $x$ such that $p(x)\neq 0$,
$\K(x \given \mathcal{S}) \leq \log(\frac{1}{p(x)})+c$.
\end{proposition}

We shall also use the following bound on conditional Kolmogorov complexity.
\begin{proposition}
 \label{prop:k-inequality}
 There is a constant $c\geq 0$ such that for any three strings $x,y,z$,
$$\K(z \given x) \geq \K(x,y) - \K(x)  - \K(y \given z,x) + \K(z \given x,y,\K(x,y))-c.$$
 \end{proposition}
\begin{proof}
Using~\cite[Theorem~3.9.1, page 232]{lv97}, 
there is a constant $c_1\geq 0$ 
such that 
$$ \abs{\K(a,b) - \K(a) - \K(b \given a,\K(a))} \leq  c_1.$$
Substituting $x,y$ for $a$ and $z$ for $b$:
$$\K(x,y) + \K(z \given x,y,\K(x,y)) - c_1 \leq \K(x,y,z) 
     \leq  \K(x) + \K(z \given x)+\K(y \given z,x) +c_2,$$
     which gives the result.
\end{proof}

\subsection{Randomized and quantum query models}
The quantum query model was implicitly introduced by Deutsch, Jozsa,
Simon and Grover~\cite{deu85,dj92,sim97,gro96}, and explicitly by
Beals, Buhrman, Cleve, Mosca and de Wolf~\cite{bbcmw01}.
In this model, as in its classical counterpart, we pay 
for accessing the oracle, but unlike the classical case,
the machine can use the power of quantum parallelism 
to make queries in superposition.
Access to the input $x\in \Sigma^n$, where $\Sigma$ is a finite set,
is achieved by way of a query operator $O_x$. The {\em query complexity} of
an algorithm is the number of calls to $O_x$.


The 
state of a computation is represented by a register $R$ composed of three subregisters:
the {\em query register} $i\in\{0,\ldots,n\}$, the {\em answer register} $z\in \Sigma$
and the {\em work register} $w$. 
We denote a register using the ket notation $\ket{R}=\ket{i}\ket{z}\ket{w}$,
or simply $\ket{i,z,w}$.
In the quantum (resp. randomized) setting, the state of the 
computation
is a complex 
(resp. non-negative real) combination of all possible values of the registers.
Let $\mathcal{H}$ denote the corresponding finite-dimensional vector space.
We denote the state of the 
computation
by a vector 
$\ket{\psi}\in\mathcal{H}$ over the basis $(\ket{i,z,w})_{i,z,w}$.
Furthermore, the state vectors are unit length
for the~$\ell_2$ norm in the quantum setting, and  for the~$\ell_1$ norm
in the randomized setting. 

A {\em $T$-query algorithm} ${A}$ is specified by a 
$(T{+}1)$-uple $(U_0,U_1,\ldots,U_T)$ of matrices.
When ${A}$ is quantum (resp., randomized),
the matrices $U_i$ are unitary (resp., stochastic).
The computation takes place as follows.
The {\em query operator} is the unitary (resp. stochastic) matrix $O_x$
that satisfies
$O_x\ket{i,z,w}=\ket{i,z\oplus x_i,w}$, for every $i,z,w$, where by
convention $x_0=0$.
Initially the state is set to some fixed value $\ket{0,0,0}$.
Then the sequence of transformations $U_0,O_x,U_1,O_x,\ldots,U_{T-1},O_x,U_T$
is applied.

We say that the algorithm ${A}$ {\em $\eps$-computes} 
a function $f:S\rightarrow S'$,
for some sets $S\subseteq\Sigma^n$ and $S'$, if the observation of the last
bits of the work register equals $f(x)$ with probability at least $1-\eps$,
for every $x\in S$.
Then $\qqc(f)$ (resp. $\rqc(f)$) is the minimum query complexity of quantum
(resp. randomized) query algorithms that $\eps_0$-compute $f$,
where $\eps_0$ is a fixed positive constant no greater than $\tfrac{1}{3}$.

\section{Proof of the main theorem}

This section is devoted to the proof of the main theorem.
We prove Theorem~\ref{thm:general}  in two main steps.  Lemma~\ref{lemma:qdivergence}
shows how fast the computations diverge when they 
start on different individual inputs, in terms of the query probabilities.
This lemma depends on the model of computation.
Lemma~\ref{lemma:queries} 
establishes the relationship between 
the Kolmogorov complexity of individual positions of the input, 
and the probability that a given algorithm makes a query to this position.
This lemma is independent
of the model of  computation.
Theorem~\ref{thm:general} follows immediately by combining these two lemmas.  

In the following two lemmas,
let $A$ be an $\eps$-bounded error algorithm for $f$ that
makes at most~$T$ queries to the input.  
Let $p^{x}_{t}(i)$ be the probability that $A$ queries $x_{i}$
at  query $t$ 
on input $x$, and let 
$\overline{p}^{x}(i)= \frac{1}{T}\sum_{t=1}^{T}p^{x}_{t}(i)$ be the 
average query probability
over all the time steps up to time $T$.
We assume henceforth without loss of generality that
$\overline{p}^{x}(i) >0$.  (For example, we start by uniformly querying 
all positions and reverse the process.)

\begin{lemma}[Divergence Lemma]~
\label{lemma:qdivergence}
For every input $x,y\in S$ such that $f(x)\neq f(y)$ the following holds.
\begin{enumerate}
\item For quantum algorithms:
$$2T\sum_{i:x_{i}\neq y_{i}}  \sqrt{\overline{p}^{x}(i)   \overline{p}^{y}(i)       }
\geq 1-2\sqrt{\eps(1-\eps)}.$$
\item For randomized algorithms:
$$2T  \sum_{i:x_{i}\neq y_{i}} \min\left(\overline{p}^{x}(i),\overline{p}^{y}(i)
\right) \geq 1-2\eps.$$
\end{enumerate}
\end{lemma}

We defer the proof of Lemma~\ref{lemma:qdivergence} to the end of 
this section.

The next lemma relates the query probabilities to the Kolmogorov complexity
of the strings.
In this lemma and the results that follow, we assume that a finite
description of the algorithm is given.  Using the knowledge of $A$, 
we may assume without loss of generality
that the function $f$ that it computes is also given, as is the length $n$
of the inputs.  With additional care, the additive constants in all of
the proofs can be made very small by adding to the auxiliary
information made available to the description algorithms, those constant-size
programs that are described within the proofs.  

\begin{lemma}[Query Information Lemma] 
\label{lemma:queries}
There exists an absolute constant $c\geq 0$ 
such that for every input $x\in S$ and position $i\in\{1,\ldots n\}$, 
$$\K(i \given x,A) \leq \log(\tfrac{1}{\overline{p}^{x}(i)})+c.$$
\end{lemma}
\begin{proof}
We describe the program that prints $i$ given $x$ and $A$.
Given $x$, use $A$ and $x$ to compute the probabilities $\overline{p}^{x}(i)$.  
This can be done in a finite number of steps because the number of queries is 
bounded by $T$.
The program includes a hard coded copy of the encoding of $i$ under
the Shannon-Fano code for this probability distribution.  Decode this
and print $i$. 
\end{proof}

{From} these two lemmas we derive the main theorem.
\begin{proof}[Proof of Theorem~\ref{thm:general}]
By Lemma~\ref{lemma:queries}, there is a constant $c\geq 0$ such that
for any algorithm that makes at most $T$ queries,
and any $x,y,i$,
$$\overline{p}^{x}(i) \leq 2^{-\K(i \given x,A)+c}
\quad\text{and}\quad
\overline{p}^{y}(i) \leq 2^{-\K(i \given y,A)+c}.$$
This is true in particular for all those $i$ where $x_i\neq y_i$.
Combining this with
Lemma~\ref{lemma:qdivergence} 
concludes the proof of the main theorem with
$C=2^{-c-1}$.
\end{proof}

We now give the proof of Lemma~\ref{lemma:qdivergence}.  The 
proof of the quantum case is very similar to the proofs found in many papers
which give quantum lower bounds on query complexity.
To our knowledge, the randomized case is new despite 
the simplicity of its proof.  
Whereas Aaronson~\cite{aar03} used a different approach to prove a version
of Ambainis'  method for randomized algorithms, our lemma
allows us to use the same framework to establish lower bounds 
for both quantum and randomized query complexities.

\begin{proof}[Proof of Lemma~\ref{lemma:qdivergence}]
Let $\ket{\psi^{x}_{t}}$ be the state of the $\eps$-bounded error algorithm $A$
just before the $t$th oracle query, on input $x$.
By convention, $\ket{\psi^{x}_{T+1}}$ is the final state.
When $A$ is a quantum algorithm $\ket{\psi^{x}_{t}}$ is a unit vector
for the~$\ell_2$-norm; otherwise it is a probabilistic distribution, that is,
a non-negative and unit vector for the~$\ell_1$-norm.
Observe that the~$\ell_1$-distance is the total variation distance.

First we prove the quantum case.  Initially, the starting state of $A$ does not
depend on the input, thus before the first question we have
$\ket{\psi^{x}_{1}} = \ket{\psi^{y}_{1}}$,
so $\braket{\psi^{x}_{1}}{\psi^{y}_{1}} = 1$.  At the end of the computation,
if the algorithm is correct with probability $\epsilon$, then
$\abs{\braket{\psi^{x}_{T+1}}{\psi^{y}_{T+1}}} \leq 2\sqrt{\epsilon(1-\epsilon)}$.
At each time step, we consider how much the two states can diverge.
\begin{claim}\label{claim1}
$\abs{\braket{\psi^{x}_{t}}{\psi^{y}_{t}}-\braket{\psi^{x}_{t+1}}{\psi^{y}_{t+1}} }
      \leq 2 \sum_{i:x_{i}\neq y_{i}} \sqrt{ p_{t}^{x}(i) p_{t}^{y}(i)}.$
\end{claim}
The proof of Claim~\ref{claim1} can be found in Appendix~\ref{app}.

Over $T$ time steps, the two states diverge as follows.  The proof uses
only Claim~\ref{claim1} and the Cauchy-Schwartz inequality.
\begin{eqnarray*}
1-2\sqrt{\eps(1-\eps)}&\leq&
\abs{ \braket{\psi^{x}_{1}}{\psi^{y}_{1}}- \braket{\psi^{x}_{T+1}}{\psi^{y}_{T+1}} }\\
 \leq \sum_{t=1}^{T}\abs{\braket{\psi^{x}_{t}}{\psi^{y}_{t}} 
                             - \braket{\psi^{x}_{t+1}}{\psi^{y}_{t+1}}}
&\leq &\sum_{t=1}^{T} 
         2 \sum_{i:x_{i}\neq y_{i}} \sqrt{ p_{t}^{x}(i) p_{t}^{y}(i)}\\
\leq 2 \sum_{i:x_{i}\neq y_{i}} \sqrt{\sum_{t=0}^{T-1} p_{t}^{x}(i)
                                                         \sum_{t=0}^{T-1}p_{t}^{y}(i)}
  &=&2T \sum_{i:x_{i}\neq y_{i}}\sqrt{\overline {p}^{x}(i)
                                                         \overline{p}^{y}(i)}.
\end{eqnarray*}

Now  we prove the randomized case. 
Again, initially $\ket{\psi^{x}_{1}} = \ket{\psi^{y}_{1}}$.
At the end of the computation,
if the algorithm is correct with probability $\epsilon$, then
$\norm{\ket{\psi^{x}_{T+1}}-\ket{\psi^{y}_{T+1}}}_1\geq 1-2\epsilon$.
At each time step, the distribution states now diverge according the following claim.
\begin{claim}\label{claim2}
$\norm{\ket{\psi^{x}_{t+1}}-\ket{\psi^{y}_{t+1}}}_1
\leq \norm{\ket{\psi^{x}_{t}}-\ket{\psi^{y}_{t}}}_1
      + 2 \sum_{i:x_{i}\neq y_{i}} \min\left(p_{t}^{x}(i),p_{t}^{y}(i)\right).
$
\end{claim}
The proof of Claim~\ref{claim2} can be found in Appendix~\ref{app}.
We now conclude the proof.
\begin{eqnarray*}
1-2\eps&\leq&
 \sum_{t=1}^{T}\norm{\ket{\psi^{x}_{t+1}}-\ket{\psi^{y}_{t+1}}}_1
-\norm{\ket{\psi^{x}_{t}}-\ket{\psi^{y}_{t}}}_1\\
&\leq&  \sum_{t=1}^{T} 
         2  \sum_{i:x_{i}\neq y_{i}} \min\left(p_{t}^{x}(i),p_{t}^{y}(i)\right)
  \leq 2T 
 \sum_{i:x_{i}\neq y_{i}} \min\left(\overline{p}^{x}(i),\overline{p}^{y}(i)\right).
\end{eqnarray*}
\end{proof}

\section{Comparison with previous adversary methods}
In this section, we reprove, as a corollary of Theorem~\ref{thm:general},
the previously known adversary lower bounds.
Our framework also allows us to obtain somewhat stronger statements for free.

To obtain the previously known adversary methods as a corollary of 
Theorem~\ref{thm:general}, we must give a lower bound
on terms $\K(i\given x,A)$ and $\K(i\given y,A)$.  To this end, we apply 
Proposition~\ref{prop:k-inequality}, and give a lower bound on $\K(x,y)$,
and upper bounds on $\K(x\given i,y)$ and $\K(y\given i,x)$.
The lower bound is obtained by applying 
Lemma~\ref{lemma:shannon-incompressibility}, a consequence of
Shannon's coding theorem, for an appropriate distribution.
The upper bounds are obtained using the Shannon-Fano code, for
appropriate distributions.

The following lemma is the general formulation of the sketch above.
\begin{lemma}\label{lemma:useful}
There exists a constant $C>0$ such that the following holds.
Let $\Sigma$ be a finite set, let $n\geq 1$ be an integer,
and let $S\subseteq\Sigma^n$. 
Let $q$ be a probability distribution on $S^2$,
let $p$ be a probability distribution on $S$
and let $\{p'_{x,i} : x\in S, 1\leq i\leq n\}$
be a set of probability
distributions on $S$.  
Then for every finite string $\sigma$,
there exist $x,y\in S$ with $q(x,y)\neq 0$, such that
$$\frac{1}{\sum_{i:x_i \neq y_i}\sqrt{2^{-\K(i\given x,\sigma)-\K(i\given y,\sigma)}}}
        \geq
       C\times \min_{i:x_i \neq y_i}
           \left(
            \frac{\sqrt{p(x)p'_{x,i}(y)\ p(y)p'_{y,i}(x)}}{q(x,y)}
           \right), \text{ and}
$$
$$
\frac{1}{\sum_{i:x_i \neq y_i} \min\left(2^{-\K(i\given x,\sigma)},
2^{-\K(i\given y,\sigma)}\right)}
\geq
C\times \min_{i:x_i \neq y_i} \left(
\max\left(
\frac{{p(x)p'_{x,i}(y)}}{q(x,y)},
\frac{{p(y)p'_{y,i}(x)}}{q(x,y)}
\right)\right),
$$
provided that $q(x,y)=0$ whenever $p(x)=0$ or $p(y)=0$, or $p'_{y,i}(x)=0$ 
or $p'_{x,i}(y)=0$ for some $i$ such that $x_i\neq y_i$.
\end{lemma}
\begin{proof}
In this proof, $c_1,\ldots,c_5$ are some appropriate nonnegative constants.
By Lemma~\ref{lemma:shannon-incompressibility},
there exists a pair $(x,y)$ such that $q(x,y)\neq 0$ and
$$\K(x,y\given \sigma,p,p') \geq \log(\tfrac{1}{q(x,y)}),$$
where $p'$ stands for a
complete description of  all the $p'_{x,i}$.

Fix $x$ and $y$ so that this holds.
By using the Shannon-Fano code (Proposition~\ref{prop:shannon}),
$$\K(x\given p)\leq\log (\tfrac{1}{p(x)})+c_1
\quad\text{and}\quad
\K(y\given x,i,p'_{x,i})\leq\log (\tfrac{1}{p'_{i,x}(y)})+c_1,$$
for any $i$ such that $x_i\neq y_i$.
By Proposition~\ref{prop:k-inequality},
\begin{eqnarray*}
\K(i\given x,\sigma) &\geq & \K(i\given x,\sigma,p,p')-c_3\\
&\geq&\K(x,y\given \sigma,p,p') - \K(x\given p)  - \K(y\given i,x,p'_{x,i}) + 
\K(i\given x,y,\K(x,y),\sigma,p,p')-c_4\\
          &\geq &  \log(\tfrac{1}{q(x,y)}) - \log(\tfrac{1}{p(x)}) 
                           -\log(\tfrac{1}{p'_{x,i}(y)}) + \K(i\given x,y,\K(x,y),\sigma,p,p')-c_5\\
          &=&\log(\tfrac{p(x)p'_{x,i}(y)}{q(x,y)}) + \K(i\given x,y,\K(x,y),\sigma,p,p')-c_5.\\
\end{eqnarray*}
Similarly,
$$\K(i\given y,\sigma) \geq \log(\tfrac{p(y)p'_{y,i}(x)}{q(x,y)})+ 
\K(i\given x,y,\K(x,y),\sigma,p,p')-c_5$$
This concludes the proof of the lemma using
Kraft's inequality (Proposition~\ref{prop:kraft}) and letting ${C=2^{-c_5}}$.
\end{proof}

\subsection{Ambainis' weighted scheme}

\begin{theorem}[Ambainis' weighted method]
\label{thm:ambainis+weights}
Let $\Sigma$ be a finite set, let $n\geq 1$ be an integer,
and let $S\subseteq\Sigma^n$ and $S'$ be sets.
Let $f:S\rightarrow S'$.
Consider a weight scheme as follows:
\begin{itemize}\setlength{\itemsep}{0pt}
\item Every pair $(x,y)\in S^2$ is assigned a non-negative weight $w(x,y)$
such that $w(x,y)=0$ whenever $f(x)=f(y)$.
\item Every triple $(x,y,i)$ is assigned
a non-negative weight $w'(x,y,i)$ such that
$w'(x,y,i)=0$ whenever $x_i=y_i$ or $f(x)=f(y)$.
\end{itemize}
For all $x,i$, let
$wt(x) {=} \sum_{y}w(x,y)$ and
$v(x,i){=} \sum_{y}w(x,y,i)$.
If $w'(x,y,i) w'(y,x,i) \geq w^{2}(x,y)$ 
for all $x,y,i$ such that $x_i \neq y_i$, 
then
$$\qqc(f)=\Omega\left(
\min_{\stackrel{x,y,i}{w(x,y)\neq 0, x_i\neq y_i}}\left(\sqrt{\frac{wt(x)wt(y)}{v(x,i)v(y,i)}}\right)
\right).$$
Furthermore, if 
$w'(x,y,i),w'(y,x,i) \geq w(x,y)$
for all $x,y,i$ such that $x_i \neq y_i$, 
then
$$\rqc(f)=\Omega\left(\min_{\stackrel{x,y,i}{w(x,y)\neq 0, x_i\neq y_i}}\left(
\max\left(\frac{wt(x)}{v(x,i)},\frac{wt(y)}{v(y,i)}
\right)\right)\right).$$
\end{theorem}
The relation in Ambainis' original statement is implicit in this
formulation, since it corresponds to the non-zero-weight pairs.
A weaker version of the randomized case was proven independently
by Aaronson~\cite{aar03} using a completely different approach.
We show that Theorem~\ref{thm:ambainis+weights} follows
from Theorem~\ref{thm:general}.  
\begin{proof}
We derive probability distributions $q,p,p'$  from the weight schemes as follows.
Let $W=\sum_{x,y}w(x,y)$ in
$$q(x,y) = \frac{w(x,y)}{W},\quad
p(x)=\frac{wt(x)}{W},
\quad\text{and}\quad
p'_{x,i}(y) =\frac{w'(y,x,i)}{v(x,i)},
\quad\text{for any $x,y,i$.}$$
It is easy to check that by construction and hypothesis, these distributions satisfy
the conditions of Lemma~\ref{lemma:useful}.  Rearranging and simplifying the
terms allows us to conclude.
\end{proof}

We conclude this section by sketching the proof
of the unweighted version of Ambainis' adversary method,
as it affords
a simpler combinatorial proof, that does not require 
Lemma~\ref{lemma:useful}.  To simplify notation we omit additive constants
and the usual auxiliary strings including $A$.  

Let $R\subseteq S\times S$, be a relation on pairs of instances,
where $(x,y)\in R {\implies} f(x) {\neq} f(y)$, and
let $R_i$ be the restriction of $R$ to pairs $x,y$ for
which $x_i \neq y_i$.
Viewing the relation $R$ as a bipartite graph, let
$l,l',m,m'$ be as follows.
\begin{itemize}\setlength{\itemsep}{0pt}
\item $m$ is a lower bound on the degree of all $x\in X$,
\item  $m'$ is a lower bound on the degree of all $y\in Y$,
\item for any fixed $x$ and $i, 1\leq i\leq n$, the number of
 $y$ adjacent to $x$ for which $x_{i}\neq y_{i}$ is at most $l$,
\item for any fixed $y$ and $i, 1\leq j\leq n$, the number of
 $x$ adjacent to $y$ for which $x_{i}\neq y_{i}$ is at most $l'$.
\end{itemize}

We make the following observations.
\begin{enumerate}
\item $\size{R} \geq \max\{m \size{X}, m'\size{Y}\}$,
so $\exists x,y~ \K(x,y)\geq \max \left(\log(m \size{X}),\log(m' \size{Y})\right).$
\item $\forall x \in X, \K(x) \leq \log(\size{X})$ and $ \K(y) \leq \log(\size{Y})$, for all $y\in Y$.
\item $\forall x,y,i \mbox{ with } (x,y) \in R_{i}, \K(y\given i,x) \leq \log(l)$ and 
similarly, $\K(x\given i,y) \leq \log(l')$.
\end{enumerate}

For any $i$ with $x_{i}\neq y_{i}$, by
Proposition~\ref{prop:k-inequality},
\begin{eqnarray*}
\K(i\given x) &\geq & \K(x,y) - \K(x)  - \K(y\given i,x) + \K(i\given x,y,\K(x,y))\\
         &\geq & \log(m\size{X})-\log(\size{X}) - \log(l) + \K(i\given x,y,\K(x,y))\\
        &=& \log(\tfrac{m}{l}) + \K(i\given x,y,\K(x,y))
\end{eqnarray*}
The same proof works to show that
$\K(i\given y) \geq \log(\frac{m'}{l'}) + \K(i\given x,y,\K(x,y)).$  
By Theorem~\ref{thm:general} and Kraft's inequality,  
$$\qqc(f) = \Omega\left(\sqrt{\tfrac{mm'}{ll'}}\right).$$

\subsection{Spectral lower bound}
We now show how to prove the spectral lower bound of
Barnum, Saks ans Szegedy~\cite{bss03} as a corollary
of Theorem~\ref{thm:general}.
Recall that for any matrix $\Gamma$, $\lambda(\Gamma)$ is the largest
eigenvalue of $\Gamma$.
\begin{theorem}[Barnum-Saks-Szegedy spectral method]\label{thm:spectra}
Let $\Sigma$ be a finite set, let $n\geq 1$ be an integer,
and let $S\subseteq\Sigma^n$ and $S'$ be sets.
Let $f:S\rightarrow S'$.
Let $\Gamma$ be an arbitrary $S\times S$ nonnegative real
symmetric matrix that satisfies
$\Gamma(x,y)=0$ whenever $f(x)=f(y)$.
For $i=1,\ldots,n$ let $\Gamma_i$ be the matrix:
$$\Gamma_i(x,y)=\begin{cases}
0,&\text{if $x_i=y_i$;}\\
\Gamma(x,y),&\text{otherwise.}
\end{cases}$$
Then $$\qqc(f)=\Omega\left(
\frac{\lambda(\Gamma)}{\max_i \lambda(\Gamma_i)}\right).$$
\end{theorem}
\begin{proof}
Let $\ket{\alpha}$ (resp. $\ket{\alpha_i}$)
be the unit eigenvector of $\Gamma$ (resp. $\Gamma_i$)
with nonnegative entries and whose eigenvalue is $\lambda(\Gamma)$
(resp. $\lambda(\Gamma_i)$).
We define the probability distributions $q,p,p'$ as follows.
Let $W=\sum_{x,y}w(x,y)$ in
$$
q(x,y) = \tfrac{\Gamma(x,y)\braket{x}{\alpha}
\braket{y}{\alpha}}{\bra{\alpha}\Gamma\ket{\alpha}},\quad
p(x)= \braket{x}{\alpha}^2,\quad
p'_{i,x}(y) =\tfrac{\Gamma_i(x,y)\braket{y}{\alpha_i}}{\bra{x}\Gamma_i\ket{\alpha_i}},
\quad\text{for any $x,y,i$}.$$

By construction these distributions satisfy
the conditions of Lemma~\ref{lemma:useful}, which suffices to conclude.
\end{proof}

\section{Certificate complexity and adversary techniques}

Let $f$ be a boolean function.
For any positive instance $x\in \Sigma^n$ of $f$ ($f(x){=}1$),
a {\em positive certificate} for $f(x)$ is the smallest subset of indices
$I\subseteq [n]$ of $x$, such that for any $y$ with
$x_i=y_i$ for all $i\in I$, $f(y){=}1$.

The {\em $1$-certificate complexity} of $f$, denoted 
$C_1(f)$, is the size of the largest positive certificate for $f(x)$, 
over all positive instances $x$.
The {\em $0$-certificate complexity}
is defined similarly for negative instances $x$ of $f$ 
($f(x)=0$).

Prior to our work, it was known that 
the best possible bound that could be proven using the unweighted
adversary technique~\cite[Theorem~5.1]{amb02} is $O(\sqrt{C_0(f)  C_1(f)})$.  
Independently,
Szegedy~\cite{sze03} showed that the best possible lower bound using the
spectral method is $O(\min(\sqrt{n C_0(f)},\sqrt{n C_1(f)}))$,
and Zhang~\cite{zha03} proved the same for Ambainis' weighted method.

The following lemma, due to Troy Lee, results in a very simple proof of
the fact that our method, and hence, all the known variants of
the adversary method, 
cannot prove lower bounds larger than
$\min(\sqrt{n C_0(f)},\sqrt{n C_1(f)}  )$.

\begin{lemma}\label{lemma:troy}
There exists a constant $c\geq 0$ such that the following holds.
Let $\Sigma$ be a finite set, let $n\geq 1$ be an integer,
and let $S\subseteq\Sigma^n$ be a set.
Let $f:S\rightarrow \{0,1\}$.
For every $x,y\in S$ with
$f(x)=0$ and   $f(y)=1$, there is an $i$ with $x_i\neq y_i$
for which $\K(i\given x,f)\leq \log(C_0(f))+c$, and similarly, there
is a $j$ with  $x_j\neq y_j$ such that $\K(j\given y,f)\leq \log(C_1(f))+c$.  
\end{lemma} 
\begin{proof}
Let $I$ be the lexicographically smallest certificate for $f(x)$.  
Since $f(x)\neq f(y)$, $x$ and~$y$ must differ on some $i\in I$. 
To describe $i$ given $x$, it suffices to give an index into $I$, which
requires $\log(C_0(f))+c$ bits.
The same can also be done with $x$ and~$y$ reversed.
\end{proof}

\begin{theorem}\label{thm:troy}
Let $\Sigma$ be a finite set, let $n\geq 1$ be an integer,
and let $S\subseteq\Sigma^n$ be a set.
Let $f:S\rightarrow \{0,1\}$.
Then any quantum query lower bound for $f$
given by Theorem~\ref{thm:general}
is in $O(\min(\sqrt{n C_0(f)},\sqrt{n C_1(f)})).$
\end{theorem}
\begin{proof}
Let $A$ be a quantum algorithm that computes $f$ with
bounded error by making at most $T$ queries to the input.
Since a description of $f$ can be obtained from a description of $A$,
$\K(i\given x,A) \leq \K(i\given x,f) + O(1)$.
Therefore, the lower bound given by Theorem~\ref{thm:general}
is $O\left(\frac{1}{\sum_{i:x_i \neq y_i}\sqrt{2^{-\K(i\given x,f)-\K(i\given y,f)}}}\right)$,
where $f(x)\neq f(y)$.
This is $O(\min(\sqrt{n C_0(f)},\sqrt{n C_1(f)} ))$
by Lemma~\ref{lemma:troy}.
\end{proof}

\section{Applications}
\subsection{A general method for distance schemes}
We generalize the technique  of H{\o}yer, Neerbek and Shi~\cite{hns02}, 
which they used to prove lower bounds on ordered search and sorting.
Though their technique is similar, 
it does not appear to be a special case of the weighted adversary method.

Here, we restrict ourselves to those weight functions that 
take values of the form $\tfrac{1}{d}$, for integer values $d$.
Therefore, instead of a weight function, we consider
an integer function $D$, which may be thought of
as a distance function on pairs of instances (even though it is not the case
in general).
We will define the {\em load} of an  instance $x$, to be the maximum 
number of instances $y$ at distance $d$ from $x$, for any $d$.
This will allow us to bound the complexity of printing $y$, given
$x$ and $d$.  (In the case of ordered search, the load will be~$1$
for all instances.)

More formally, for any non-negative integer function $D$ on pairs $(x,y)$, we
define the {\em right load}  $\rightload(x,i)$ to be the
maximum over all values $d$, of the 
number of $y$ such that $D(x,y) = d$ and $x_i\neq y_i$.
The {\em left load} $\leftload(y,i)$ is defined similarly, 
inverting $x$ and $y$.
\begin{theorem}
\label{thm:hoyer}
Let $\Sigma$ be a finite set, let $n\geq 1$ be an integer,
and let $S\subseteq\Sigma^n$ and $S'$ be sets.
Let $f:S\rightarrow S'$.
Let $D$ be a non-negative integer function on $S^2$
such that $D(x,y)=0$ whenever $f(x)=f(y)$.
Let $W=\sum_{x,y : D(x,y)\neq 0}\tfrac{1}{D(x,y)}$.
Then
$$\qqc(f)=\Omega\left(\frac{W}{\size{S}}\min_{\stackrel{x,y}{D(x,y)\neq 0, x_i=y_i}}
\left(
\frac{1}{\sqrt{ \rightload(x,i)\leftload(y,i)}}\right)\right),$$
$$
\rqc(f)=\Omega\left(\frac{W}{\size{S}}\min_{\stackrel{x,y}{D(x,y)\neq 0, x_i=y_i}}\left(
\max\left(
\frac{1}{ \rightload(x,i)},
\frac{1}{ \leftload(y,i)}\right)\right)\right).
$$
\end{theorem}
\begin{proof}
We use a variation on Lemma~\ref{lemma:useful}.
We define probability
distributions $q(x,y)=\tfrac{1}{D(x,y)\times W}$ whenever
$D(x,y)\neq 0$ and $q(x,y)=0$ otherwise; 
$p(x)=\tfrac{1}{\size{S}}$.
Fix $\sigma$ to be the string containing a description of $A$ and $D$, where
$D$ is a complete description of the 
distance function, and where we assume that $A$ includes a description of 
$f$, hence $\rightload(x,i), \leftload(y,i)$ and are  also given.

We give an upper bound on the terms $\K(y\given x,i)$ and $\K(x\given y,i)$
directly, using left and right loads.
Given $x,i$ and some integer $d>0$, there are
at most $\rightload(x,i)$ instances $y$ such that
$D(x,y)=d$ and $x_i\neq y_i$. Therefore
$$\K(y\given x,i,\sigma)\leq \log (D(x,y))+ \log(\rightload(x,i))+c,$$
where $c\geq 0$ is some constant, 
The same is true for $\K(x\given y,i)$:
$$\K(y\given x,i,\sigma)\leq \log (D(x,y))+ \log(\leftload(y,i))+c.$$

Now, we conclude
following the same sketch as the proof of Lemma~\ref{lemma:useful}.
\end{proof}

We reprove some of the lower bounds of H{\o}yer, Neerbek and Shi.
The distance schemes we use are exactly the ones of~\cite{hns02}.
Whereas they did not separate the quantum part from the combinatorial part in
their proofs, here we only need to evaluate the combinatorial objects 
$\rightload$ and $\leftload$ to get the results.
\begin{corollary}
$\qqc(\textsc{Ordered search}),\rqc(\textsc{Ordered search})=\Omega(\log n).$
\end{corollary}
\begin{proof}
Fix $\Sigma=\{0,1\}$.
We only consider the set of instances $S$ of length $n$ 
of the form $0^{a-1}1^{n-a}$. Note that $\size{S}=n$.
Define distance for pairs $(x,y)\in S^2$ as
$D(x,y)={b-a}$, and  $D(x,y)=0$ for all other instances, where
$x=0^{a-1}1^{n-a}$ and $y=0^{b-1}1^{n-b}$
with $1\leq a<b\leq n$. 
The inverse distance has total weight $W=\Theta(n\log n)$.
Furthermore, for every $x,y,i$ such that $D(x,y)\neq 0$ and $x_i\neq y_i$,
$\rightload(x,i)=\leftload(y,i)=1$.
The result follows by Theorem~\ref{thm:hoyer}.
\end{proof}

A lower bound for sorting~\cite{hns02} in the
comparison model can also be obtained
by applying Theorem~\ref{thm:hoyer}.
\begin{corollary}
$\qqc(\textsc{Sorting}),\rqc(\textsc{Sorting})=\Omega(n\log n).$
\end{corollary}
\begin{proof}
Fix $\Sigma=\{0,1\}$.
An input is an $n\times n$ comparison matrix $M_\sigma$
defined by $(M_\sigma)_{i,j}=1$ if $\sigma(i)<\sigma(j)$,
and $(M_\sigma)_{i,j}=0$ otherwise, where $\sigma$ is some
permutation of $\{1,\ldots,n\}$.
(In the usual array representation, the element of rank $r$ in the
array would be stored at position $\sigma^{-1}(r)$.)
The set $S$ of inputs is $\{M_\sigma:\sigma \in S_n\}$.

We consider pairs of instances $M_\sigma, M_{\sigma^{(k,d)}}$,
where $\sigma^{(k,d)}$ is obtained from $\sigma$ by changing the
value of the element of rank $k+d$  to a value that immediately precedes
the element of rank $k$ in $\sigma$.  This changes the rank of the $d$
elements of intermediate rank, incrementing their rank by one.

More formally, define $\sigma^{(k,d)}=(k,k+1,\ldots,k+d)\circ\sigma$,
for $d\neq 0$.
For every permutations $\sigma,\tau$ we let
$D(M_\sigma,M_\tau)={d}$ if there exists $k,d$
such that $\tau=\sigma^{(k,d)}$, and $D(M_\sigma,M_\tau)=0$
otherwise.
Observe that whenever $\tau=\sigma^{(k,d)}$, the
comparison matrices $M_\sigma$ and $M_\tau$ differ only
in entries $(\sigma^{-1}(k+d),\sigma^{-1}(i))=(\tau^{-1}(k),\tau^{-1}(i+1))$
and $(\sigma^{-1}(i),\sigma^{-1}(k+d))=(\tau^{-1}(i+1),\tau^{-1}(k))$,
for $k\leq i\leq k+d-1$.

Then for every $\sigma,\tau,(i,j)$ such that $D(M_\sigma,M_\tau)\neq 0$
and $(M_{\sigma})_{i,j}\neq (M_\tau)_{i,j}$, 
$\rightload(\sigma,(i,j))=\leftload(\sigma,(i,j))=2$.
This is because given $\sigma,i,j,d$, 
either $i=\sigma^{-1}(k+d)$ or $j=\sigma^{-1}(k+d)$,
so there are two possible values for $(k,d)$.
Similarly, $\rightload(\tau,(i,j))=\leftload(\tau,(i,j))=2$.
The inverse distance has total weight $W=\Theta((n!) n\log n)$
and the size of $S$ is $\size{S}=(n!)$.
Applying Theorem~\ref{thm:hoyer}, we conclude the proof.
\end{proof}

\subsection{Graph properties}

Theorem~\ref{thm:general} provides a simple and intuitive method to
prove lower bounds for specific problems.  We illustrate this by
giving lower bounds for two graph properties: connectivity, and
bipartiteness.  These are direct applications of Theorem~\ref{thm:general}
in that we analyze directly the complexity $\K(i|x,A)$ without defining
relations or weights or distributions: we only need to consider
a ``typical'' hard pair of instances.
In this section, we omit additive and multiplicative 
constants that result from using small, constant-size programs,
as well as the constant length auxiliary string $A$
to simplify the proofs.

\subsubsection{Bipartiteness}

\begin{theorem}[\cite{dhhml03}]\label{thm:bipart}
In the adjacency matrix model, 
$$\qqc(\textsc{Bipartiteness})=\Omega(n),$$
where $n$ is the number of vertices in the graph.
\end{theorem}
\begin{proof}
Let $G$ be the star on $n$ vertices.  Let $i,j$ be two leaves of $G$
chosen such that $\K(i,j|G) \geq  \log{n \choose 2}$.
Define $H$ to be $G$ to which the single edge $(i,j)$ is added.

\begin{figure}[h]
\begin{center}
\includegraphics[height=3cm]{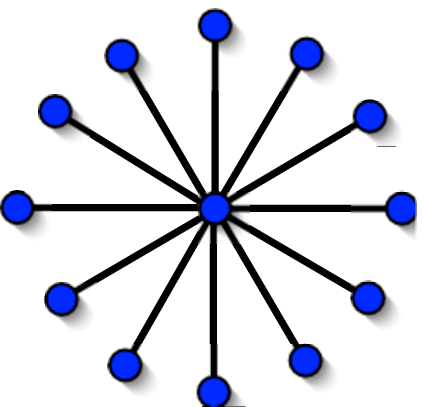}
\quad
\includegraphics[height=3cm]{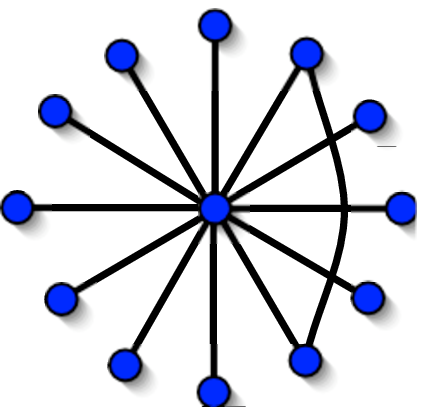}
\quad
\caption{ 
Graphs G, H for the bipartiteness lower bound
}
\end{center}
\end{figure}
By hypothesis, $\K(i,j|G) \geq \log{n \choose 2}$. 
By Theorem~\ref{thm:general}, 
$\qqc(\textsc{Bipartiteness})=\Omega(n),$ as claimed.
\end{proof}

\subsubsection{Graph connectivity}
\begin{theorem}[\cite{dhhml03}]\label{thm:connect}
In the adjacency matrix model, 
$$\qqc(\textsc{GraphConnectivity})=\Omega(n^{3/2}),$$
where $n$ is the number of vertices in the graph.
\end{theorem}
\begin{proof}
It suffices to consider one negative and one positive instance
of graph connectivity, which we construct using the incompressibility method,
using the ideas of~\cite{dhhml03}.
Let $S$ be an incompressible string of length $\log(n-1)!+\log {n\choose 2} $,
chopped into two pieces $S_1$ and $S_2$ of length $\log(n-1)!$ 
and $\log{n\choose 2}$, respectively.  We think of $S_1$ as representing
a hamilton cycle $C=(0, \pi(0) \cdots \pi(n-1), 0)$ through the $n$ vertices, 
and $S_2$ as representing a pair of distinct vertices $s,t$.  
Let $G$ contain the cycle $C$ and let $H$ be obtained from $G$ by breaking
the cycle into two cycles at $s$ and $t$, that is, 
$H = G \setminus \{(\pi(s),\pi(s+1)), 
(\pi(t),\pi(t+1))\} \cup \{(\pi(s), \pi(t+1)), (\pi(s+1),\pi(t))\}$.

\begin{figure}[h]
\begin{center}
\includegraphics[height=4cm]{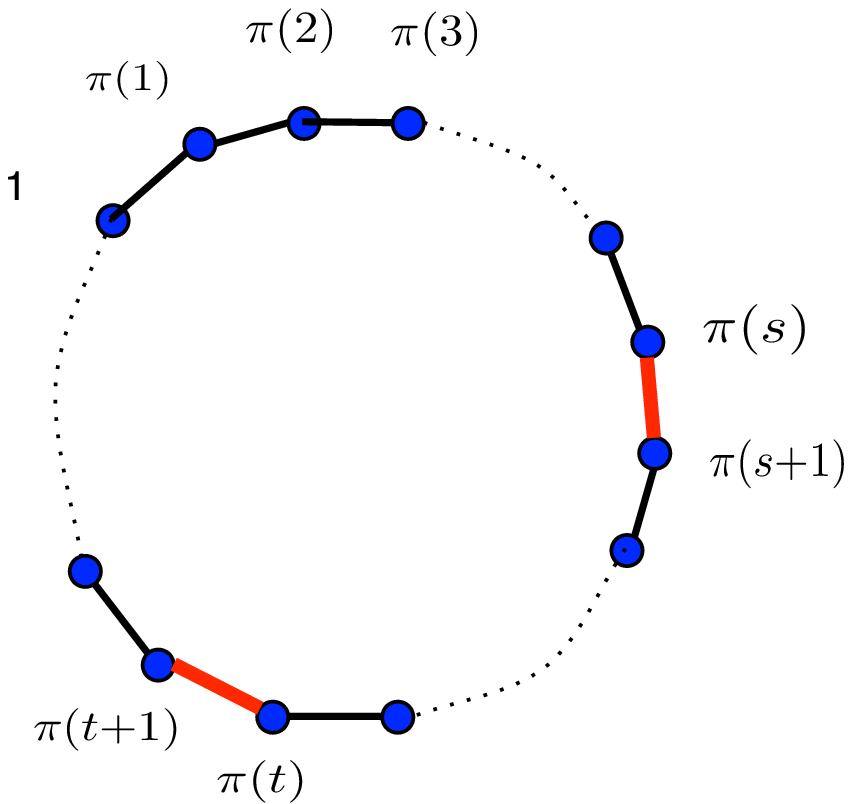}
\quad
\includegraphics[height=4cm]{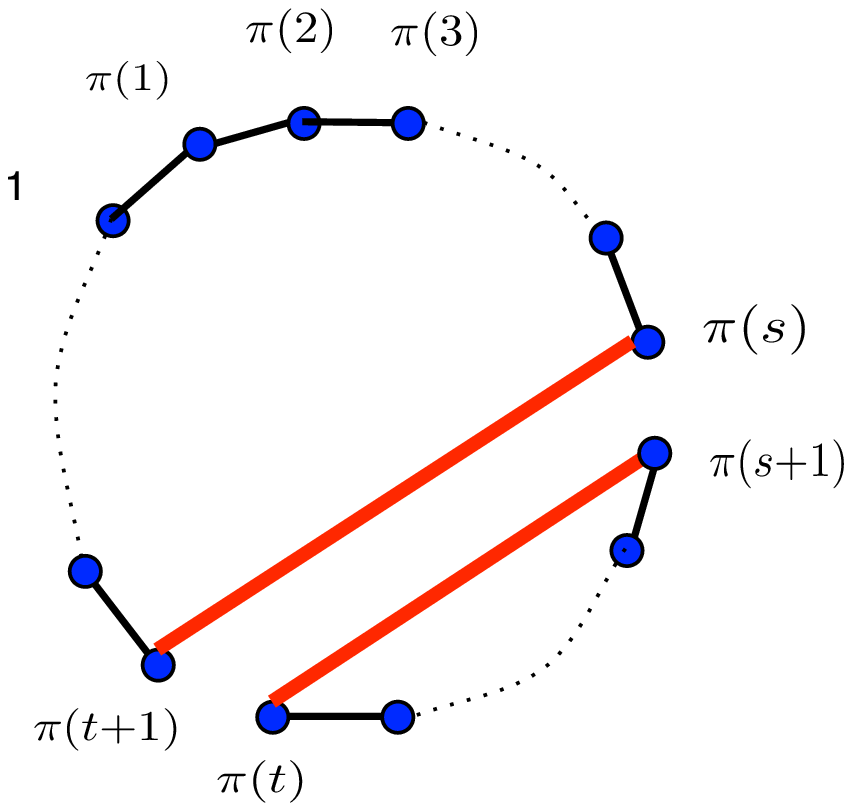}
\quad
\caption{ 
Graphs G, H for the connectivity lower bound
}
\end{center}
\end{figure}

We show that for the four edges $e$ where $G$ and $H$ differ, 
$\K(e\given G)+\K(e\given H) \geq 3\log n  - 4.$  
Let $e_-, e_-'$ be the edges removed from $G$, and
$e_+, e_+'$ be the edges added to $G$.
Observe that up to an additive constant, 
$\K(e_+ \given G)=\K(e_+'\given G)$ and $\K(e_- \given H)=\K(e_-'\given H)$.
Let $e_{-}$ be one of the edges removed from
$G$, w.l.o.g., $e_{-}=(\pi(s),\pi(s+1))$.
\begin{eqnarray*}
\log(n-1)!+\log{n\choose 2} \leq \K(S) &\leq&  \K(G) + \K(s\given G) + \K(t\given G)\\
 & \leq & \K(G) + \K(e_{-}\given G) + \log n \\
\K(e_{-}\given G) &\geq & \log{n\choose 2}- \log n = \log\tfrac{n-1}{2}\\ 
\end{eqnarray*}

Assume w.l.o.g.  that the smallest cycle of $H$ 
contains $\pi(s)$, and let $l$ be its length.
\begin{eqnarray*}
\log(n-1)!+\log{n\choose 2} \leq \K(S) 
       &\leq & \K(H) + \K(e_{-}\given H) + \K(\pi(t),\pi(t+1)\given H)\\
       & \leq & \log\tfrac{(n-1)!}{(n-l+1)!} + \log(n{-}l{-}1)!
                        + \K(e_{-}\given H)  + \log l + \log(n{-}l)\\
\K(e_{-}\given H) &\geq & 2 \log n +  \log(n{-}l) -\log(l) \geq  2\log n .\\
\end{eqnarray*}

For the added edges, $e_{+}, e_{+}'$, consider w.l.o.g.
$e_{+}=(\pi(s), \pi(t+1))$. Since $S$ is incompressible,
        $\K(e_{+}\given G) \geq \K(s,t\given G) \geq\log{n \choose 2}$.
Furthermore, $\K(S) \leq \K(H) + \K(e_{+}\given H) + \K(e_{+}'\given H)$, 
and $\K(e_{+}'\given H) \leq \log n $, so 
$\K(e_{+}\given H) \geq \log{n\choose 2} - \log n  = \log\tfrac{n-1}{2} .$
The same proof shows that $\K(e_{+}'\given H) \geq \log \tfrac{n-1}{2} .$

\end{proof}

\section{Acknowledgements}
We wish to thank Troy Lee, Christoph D\"urr for many useful discussions,
and Andris Ambainis for his helpful answers to our questions.

\bibliographystyle{alpha}
\bibliography{lm03}

\appendix

\section{Proofs of claims}
\label{app}
\begin{proof}[Proof of Claim~\ref{claim1}]
Let
$$\ket{\psi^{x}_{t}} =  \sum_{i,z,w}\alpha_{i,z,w}\ket{i,z,w},
\quad\text{and}\quad
\ket{\psi^{y}_{t}} = \sum_{i,z,w}\beta_{i,z,w}\ket{i,z,w}.$$
 After the $t$th query is made,  the states $\ket{\psi'^{x}_{t}}=O_x\ket{\psi^{x}_{t}}$
and $\ket{\psi'^{y}_{t}}=O_y\ket{\psi^{y}_{t}}$ are
$$\ket{\psi'^{x}_{t}} = \sum_{i,z,w}\alpha_{i,z,w}\ket{i,z\oplus x_i,w},
\quad\text{and}\quad
\ket{\psi'^{y}_{t}} = \sum_{i,z,w}\beta_{i,z,w}\ket{i,z\oplus y_i,w}.$$
 
Now, since the inner product is invariant under unitary transformations,
we get 
$$\braket{\psi^{x}_{t+1}}{\psi^{y}_{t+1}}=
\braket{\psi'^{x}_{t}}{\psi'^{y}_{t}},$$
and therefore,
\begin{eqnarray*}
\abs{\braket{\psi^{x}_{t}}{\psi^{y}_{t}} - \braket{\psi^{x}_{t+1}}{\psi^{y}_{t+1}}}
   &=&   \abs{\sum_{i,z,w}\overline{\alpha_{i,z,w} }\beta_{i,z,w}
         - \sum_{i,z,w}\overline{\alpha_{i,z\oplus x_{i},w} }\beta_{i,z\oplus y_{i},w}}\\
    &=&  \abs{\sum_{\stackrel{i,z,w}{x_{i}\neq y_{i}}}\overline{\alpha_{i,z,w} }\beta_{i,z,w}
         -               \overline{\alpha_{i,z\oplus x_{i},w} }\beta_{i,z\oplus y_{i},w}}\\
     & \leq &
\sum_{i:x_{i}\neq y_{i}}\left(
\abs{\sum_{z,w}
                      \overline{\alpha_{i,z,w} }\beta_{i,z,w}}
         + \abs{\sum_{z,w}
                       \overline{\alpha_{i,z\oplus x_{i},w} }\beta_{i,z\oplus y_i,w}} 
					   \right)\\
    & \leq & 
2\sum_{i:x_{i}\neq y_{i}}
           \sqrt{\left( \sum_{z,w}
                                  \abs{{\alpha_{i,z,w} }}^{2}
                     \right)
                     \left( \sum_{z,w}
                                  \abs{{\beta_{i,z,w} }}^{2}
                     \right)}
					\\
     &\leq& 2 \sum_{i:x_{i}\neq y_{i}} \sqrt{ p_{t}^{x}(i) p_{t}^{y}(i)}
\end{eqnarray*}
\end{proof}

\begin{proof}[Proof of Claim~\ref{claim2}]
Let us write the distributions using the above formalism, that is,
$$\ket{\psi^{x}_{t}} =  \sum_{i,z,w}\alpha_{i,z,w}\ket{i,z,w},
\quad\text{and}\quad
\ket{\psi^{y}_{t}} = \sum_{i,z,w}\beta_{i,z,w}\ket{i,z,w}.$$
 Note that now, the vectors are unit for the~$\ell_1$ norm.
 After the $t$th query is made,  the states $\ket{\psi'^{x}_{t}}=O_x\ket{\psi^{x}_{t}}$
and $\ket{\psi'^{y}_{t}}=O_y\ket{\psi^{y}_{t}}$ are
$$\ket{\psi'^{x}_{t}} = \sum_{i,z,w}\alpha_{i,z,w}\ket{i,z\oplus x_i,w},
\quad\text{and}\quad
\ket{\psi'^{y}_{t}} = \sum_{i,z,w}\beta_{i,z,w}\ket{i,z\oplus y_i,w}.$$
 
Now, since the~$\ell_1$ distance does not increase under stochastic matrices,
we get 
$$\norm{\ket{\psi^{x}_{t+1}}-\ket{\psi^{y}_{t+1}}}_1
\leq
\norm{\ket{\psi'^{x}_{t}}-\ket{\psi'^{y}_{t}}}_1,$$
 and therefore,
\begin{eqnarray*}
\norm{\ket{\psi^{x}_{t+1}}-\ket{\psi^{y}_{t+1}}}_1
&=&\norm{\sum_{i,z,w}(\alpha_{i,z,w}\ket{i,z\oplus x_i,w}-
\beta_{i,z,w}\ket{i,z\oplus y_i,w})}_1\\
&=&\sum_i\norm{\sum_{z,w}(\alpha_{i,z,w}\ket{i,z\oplus x_i,w}-
\beta_{i,z,w}\ket{i,z\oplus y_i,w})}_1.
\end{eqnarray*}
We now bound each term of the last sum separately. Fix any $i$.
If $x_i=y_i$ then
$$\norm{\sum_{z,w}(\alpha_{i,z,w}\ket{i,z\oplus x_i,w}-
\beta_{i,z,w}\ket{i,z\oplus y_i,w})}_1=
\norm{\sum_{z,w}(\alpha_{i,z,w}\ket{i,z,w}-
\beta_{i,z,w}\ket{i,z,w})}_1.$$
If $x_i\neq y_i$ then,
\begin{eqnarray*}
&&\norm{\sum_{z,w}(\alpha_{i,z,w}\ket{i,z\oplus x_i,w}-
\beta_{i,z,w}\ket{i,z\oplus y_i,w})}_1\\
&\leq&\norm{\sum_{z,w}(\alpha_{i,z,w}\ket{i,z\oplus y_i,w}-
\beta_{i,z,w}\ket{i,z\oplus y_i,w})}_1\\
&&+\norm{\sum_{z,w}(\alpha_{i,z,w}\ket{i,z\oplus x_i,w}-
\alpha_{i,z,w}\ket{i,z\oplus y_i,w})}_1\\
&\leq& 
\norm{\sum_{z,w}(\alpha_{i,z,w}\ket{i,z,w}-
\beta_{i,z,w}\ket{i,z,w})}_1
+2
\norm{\sum_{z,w}\alpha_{i,z,w}\ket{i,z,w}}_1\\
&=&\norm{\sum_{z,w}(\alpha_{i,z,w}\ket{i,z,w}-
\beta_{i,z,w}\ket{i,z,w})}_1
      + 2  p_{t}^{x}(i).
\end{eqnarray*}
In the same way we can prove that
\begin{eqnarray*}
&&\norm{\sum_{z,w}(\alpha_{i,z,w}\ket{i,z\oplus x_i,w}-
\beta_{i,z,w}\ket{i,z\oplus y_i,w})}_1\\
&\leq&\norm{\sum_{z,w}(\alpha_{i,z,w}\ket{i,z,w}-
\beta_{i,z,w}\ket{i,z,w})}_1
      + 2  p_{t}^{y}(i).
\end{eqnarray*}
We regroup these majorations and conclude
\begin{eqnarray*}
\norm{\ket{\psi^{x}_{t+1}}-\ket{\psi^{y}_{t+1}}}_1
&\leq&
\sum_i\norm{\sum_{z,w}(\alpha_{i,z,w}\ket{i,z,w}-
\beta_{i,z,w}\ket{i,z,w})}_1
+2\sum_{i:x_i\neq y_i}\min\left(p_{t}^{x}(i),p_{t}^{y}(i)\right)\\
&=&\norm{\ket{\psi^{x}_{t}}-\ket{\psi^{y}_{t}}}_1
+2\sum_{i:x_i\neq y_i}\min\left(p_{t}^{x}(i),p_{t}^{y}(i)\right).
\end{eqnarray*}

\end{proof}

\end{document}